\documentclass[useAMS,usenatbib]{mn2e}
\usepackage{amsmath}
\usepackage{epsfig}
\usepackage{amssymb}
\usepackage{graphicx}
\usepackage{times}

\begin{document}

\title[Models for four galaxies in the Ursa Major cluster]
{Finite thin disc models of four galaxies in the Ursa Major cluster: NGC3877,
NGC3917, NGC3949 and NGC4010}

\author[G. A. Gonz\'alez, S. M. Plata-Plata and J. Ramos-Caro]
{Guillermo A. Gonz\'alez$^{1,2}$\thanks{E-mail: guillego@uis.edu.co (GAG);
sandra.plata@gmail.com (SMP-P); javiramos1976@gmail.com (JR-C)},
Sandra M. Plata-Plata$^{1}$\footnotemark[1] and
Javier Ramos-Caro$^{1}$\footnotemark[1] \\
$^{1}$Escuela de F\'isica, Universidad Industrial de Santander, A. A. 678,
Bucaramanga, Colombia\\
$^{2}$Departamento de F\'isica Te\'orica, Universidad del Pa\'is Vasco, 48080
Bilbao, Spain}

\maketitle

\begin{abstract}
Finite thin disc models of four galaxies in the Ursa Major cluster are
presented. The models are obtained by means of the Hunter method and the
particular solutions are choosen in such a way that the circular velocities are
adjusted very accurately to the observed rotation curves of some specific spiral
galaxies. We present particular models for the four galaxies NGC3877, NGC3917,
NGC3949 and NGC4010 with data taken from the recent paper by Verheijen \&
Sancici (2001). By integrating the corresponding surface mass densities, we
obtain the total mass $\mathcal M$ of these four galaxies, all of them being of
the order of $10^{10} \mathcal{M_{\odot}}$. These obtained values for $\mathcal
M$ may be taken as a quite accurately estimative of the mass upper bound of
these galaxies, since in the model was considered that all their mass was
concentrated at the galactic disc. The models can be consider as a first
approximation to the obtaining of quite realistic models of spiral galaxies. 
\end{abstract}


\section{Introduction}

Currently, the most accepted description of the composition of spiral galaxies
is that a main part of its mass is concentrated in a thin disc, being the other
constituents a spheroidal halo, a central bulge and, perhaps, a central black
hole (\citeauthor{BT} \citeyear{BT}). Now, as all of these components contribute
to the gravitational field of a galaxy, obtaining proper models that include the
effects of all parts is a rather difficult problem. However, the contribution of
all parts is limited to certain distance scales, so that not all components have
to be included in a reasonably realistic model. Therefore, it is commonly
accepted that many of the main aspects of the galactic dynamics can be
described, in a quite approximate way, with models that only consider the thin
galactic disc.

Accordingly, the study of the gravitational potential generated by an idealized
thin disc is a problem of great astrophysical relevance and so, through the
years, different approaches have been used to obtain such kind of thin disc
models (see \cite{BT} and references therein). So, once an expression for the
gravitational potential has been derived, corresponding expressions for the
surface mass density of the disc and for the circular velocity of the disc
particles can be obtained. Then, if the expression for the circular velocity can
be adjusted to fit the observational data of the rotation curve of a particular
galaxy, the total mass can be obtained by integrating the corresponding surface
mass density.

However, although most of these thin disc models have surface densities and
rotation curves with remarkable properties, many of them mainly represent discs
of infinite extension and thus they are rather poor flat galaxy models.
Therefore, in order to obtain more realistic  models of flat galaxies, is better
to consider methods that permit the obtention of finite thin disc models. Now, a
simple method to obtain the gravitational potential, the surface density and the
rotation curve of thin discs of finite radius was developed by \cite{HUN1}, the
simplest example of a disc obtained by this method being the well known
\cite{KAL} disc.

In a previous paper (\citeauthor{GR} \citeyear{GR}) we use the Hunter method in
order to obtain an infinite family of thin discs of finite radius with a well
behaved surface mass density, an infinite family of generalized Kalnajs discs.
Also, the motion of test particles in the gravitational fields generated by the
first four members of this family was studied (\citeauthor{RLG} \citeyear{RLG})
and a new infinite family of self-consistent models was obtained as a
superposition of members belonging to the generalized Kalnajs family
(\citeauthor{PRG} \citeyear{PRG}).

In \cite{GR}, the family of disc models was derived by requiring that the
surface density behaves as a monotonously decreasing function of the radius,
with a maximum at the center of the disc and vanishing at the edge. So, although
the mass distribution of this family of discs present a satisfactory behavior in
such a way that they could be considered adequate as flat galaxy models, their
corresponding rotation curves do not present a so good behavior as they do not
reproduce the flat region of the observed rotation curve.

On the other hand, in \cite{PRG} the new family of discs was obtained by
superposing the members of the generalized Kalnajs family in order that the
resulting surface density can be expressed as a well behaved function of the
gravitational potential, in such a way that the corresponding distribution
functions can be easily obtained. Furthermore, besides present a well-behaved
surface density, the models also present rotation curves with a better behavior
than the generalized Kalnajs discs and are radially stable, whereas vertically
unstable. Then, apart of the stability problems, these discs can be considered
as quite adequate models in order to describe satisfactorily a great variety of
galaxies.

In agreement with the above considerations, in this paper we explore the
possibility of obtain some thin disc models in which the circular velocities can
be adjusted very accurately to fit the observed rotation curves. In order to do
this, we will consider a different approach in order to solve the boundary value
problem defining the gravitational potential. So, instead of assuming a given
behavior of the surface density, we will obtain from the general solution of the
Laplace equation a quite general expression for the circular velocity, which it
is expressed as a power series expansion of a dimensionless radial coordinate.
Then, after determining the coefficients of the series expansion by means of a
proper numerical adjust, the corresponding surface densities and all the
quantities characterizing the kinematic behavior of the  particular models can
be obtained. Also, the total mass of the galaxies can be obtained by integrating
their surface density.

The paper is organized as follows. First, in Section \ref{sec:gen}, closely
following \cite{GR}, we present the general finite thin disc model obtained by
means of the Hunter method. Then, in Section \ref{sec:fam}, we particularize the
model by taking a finite number of terms in the series expansion of the general
model. Also we obtain explicit expressions for the kinematical quantities
characterizing the behavior of the discs. In Section \ref{sec:fit} the models
are then fitted to data of the observed rotation curves of four galaxies in the
Ursa Major cluster, as reported in \cite{VS}. Finally, in Section
\ref{sec:disc}, we discuss the obtained results.

\section{General Finite Thin Disc Models}\label{sec:gen}

We begin by considering the Laplace equation for an axially symmetric problem,
written as
\begin{equation}
\frac{\partial^2 \Phi}{\partial R^2} + \frac{1}{R} \frac{\partial \Phi
}{\partial R}+ \frac{\partial^2 \Phi}{\partial z^2} = 0,
\end{equation}
where $(R,\varphi,z)$ are the usual cylindrical coordinates. We will suppose
also that, besides the axial symmetry, the gravitational potential has symmetry
of reflection with respect to the plane $z = 0$,
\begin{equation}
\Phi(R,z) = \Phi(R,-z),
\end{equation}
which implies that the normal derivative of the potential satisfies the relation
\begin{equation}
\frac{\partial \Phi}{\partial z}(R,-z) = - \frac{\partial \Phi}{\partial
z}(R,z), \label{simetriarefl}
\end{equation}
in agreement with the attractive character of the gravitational field.  We also
assume that $\partial \Phi / \partial z$ does not vanish on the plane $z = 0$,
in order to have a thin distribution of matter that represents the disc.

So, given a potential $\Phi(R,z)$ with the previous properties, we can easily
obtain the circular velocity $v_c (R)$, defined as the tangential velocity of
the stars in circular orbits around the center, through the relation
\begin{equation}
v_{c}^{2}(R) =  R \left. \frac{\partial \Phi}{\partial R} \right|_{z=0}.
\label{vc2} 
\end{equation}
Also, given $\Phi(R,z)$, the density $\Sigma(R)$ of the surface distribution of
matter can be obtained using the Gauss law and, by using the equation
(\ref{simetriarefl}), we obtain
\begin{equation}
\Sigma(R) = \left. \frac{1}{2 \pi G} \frac{\partial \Phi}{\partial
z} \right|_{z=0^{+}}.
\end{equation}
Thus, in order to have a surface density corresponding to a finite disclike distribution
of matter, we impose boundary conditions in the form
\begin{subequations}
\begin{align}
\left. \frac{\partial \Phi}{\partial z} \right|_{z=0^{+}} \neq 0; \quad R \leq
a, \\
\left. \frac{\partial \Phi}{\partial z} \right|_{z=0^{+}} = 0; \quad R > a,
\end{align}
\end{subequations}
in such a way that the matter distribution is restricted to the disc $z = 0$, $0
\leq R \leq a$.

In order to properly pose the boundary value problem, we introduce oblate
spheroidal coordinates, whose symmetry adapts in a natural way to the geometry
of the model. These coordinates are related to the usual cylindrical coordinates
by the relation (\citeauthor{MF} \citeyear{MF})
\begin{subequations}
\begin{align}
R &= a \sqrt{(1+\xi^{2})(1-\eta^{2})},\\
z &= a \xi \eta,
\end{align}
\end{subequations}
where $0 \leq \xi < \infty$ and $-1 \leq \eta < 1$.  The disc has the
coordinates $\xi=0$, $0 \leq \eta^{2} < 1$.  On crossing the disc, the $\eta$
coordinate changes sign but does not change in absolute value.  The singular
behaviour of this coordinate implies that an even function of $\eta$ is a
continuous function everywhere but has a discontinuous $\eta$ derivative at the
disc.

In terms of the oblate spheroidal coordinates, the Laplace equation can be
written as
\begin{equation}
\frac{\partial }{\partial \xi} \left[(1 + \xi^{2}) \frac{\partial \Phi}{\partial
\xi} \right] + \frac{\partial }{\partial \eta} \left[(1 - \eta^{2})
\frac{\partial \Phi}{\partial \eta} \right] = 0,
\end{equation}
and we need to find solutions that are even functions of $\eta$ and with
the boundary conditions
\begin{subequations}
\begin{align}
\left. \frac{\partial \Phi}{\partial \xi} \right|_{\xi = 0} &\neq 0, \\
\left. \frac{\partial \Phi}{\partial \eta} \right|_{\eta = 0} &= 0.
\end{align}
\end{subequations}
According to this, the Newtonian gravitational potential for the exterior of a
finite thin disc with an axially symmetric matter density can be written as
(\citeauthor{BAT} \citeyear{BAT}),
\begin{equation}
\Phi(\xi,\eta) = - \sum_{n=0}^{\infty} C_{2n} q_{2n}(\xi) P_{2n}(\eta),
\label{eq:poten}
\end{equation}
where $C_{2n}$ are arbitrary constants, $P_{2n}(\eta)$ and $q_{2n}(\xi)=
i^{2n+1}Q_{2n}(i\xi)$ are the usual Legendre polynomials and the Legendre
functions of second kind, respectively.

With this general solution for the gravitational potential, the circular
velocity can be written as
\begin{equation}
v_{c}^{2}({\widetilde R}) = \frac{{\widetilde R}^2}{\eta} \sum_{n = 1}^{\infty} C_{2n} q_{2n}(0)
P'_{2n}(\eta), \label{velocity}
\end{equation}
while the surface matter density is given by
\begin{equation}
\Sigma({\widetilde R}) = \frac{1}{2\pi a G \eta}\sum_{n = 0}^{\infty} C_{2n} (2n+1)
q_{2n+1}(0) P_{2n}(\eta), \label{density}
\end{equation}
where $\eta = \sqrt{1 - {\widetilde R}^2}$ and ${\widetilde R} = r/a$. So, by
integrating on the total area of the disc, we find the expression
\begin{equation}
\frac{\mathcal{M} G}{a} = C_0, \label{masa}
\end{equation}
which allows to compute the value of the total mass $\mathcal{M}$. Accordingly, all the
quantities characterizing the thin disc model are determined in terms of the set
of constants $C_{2n}$, which can be determined from the observational data
corresponding to rotation curves for some particular galaxy.

\section{A Family of Particular Models}\label{sec:fam}

We now explore the possibility of obtain particular thin disc models in which
the expression for the circular velocity can be very accurately adjusted with
the observed data from the rotation curve of a given galaxy. However, in order
to do this, first the sum must be limited to a finite number of terms. This
correspond to take $C_{2n} = 0$ for $n > m$, with $m$ a positive integer. So,
after replace the derivatives of the Legendre polynomials, the expression
(\ref{velocity}) can be cast as
\begin{equation}
v_{c}^{2}({\widetilde R}) = \sum_{n = 1}^{m} A_{2n} {\widetilde R}^{2n},
\label{velr2n}
\end{equation}
where the $A_{2n}$ constants are related with the previous constants $C_{2n}$
through the relation
\begin{equation}
C_{2n}=\frac{4n + 1}{4n(2n + 1)}\sum_{k = 1}^{m} \frac{A_{2k}}{q_{2n}(0)}
\int_{-1}^{1} \eta(1 - \eta^{2})^{k} P'_{2n}(\eta) d\eta, \label{relation}
\end{equation}
for $n \neq 0$, which is obtained by equaling expressions (\ref{velocity}) and
(\ref{velr2n}) and by using properties of the Legendre polynomials
(\citeauthor{AW} \citeyear{AW}). Then, if the constants $A_{2n}$ are determined
by a fitting of the observational data of the corresponding rotation curve, the
relation  (\ref{relation}) gives the values of the constants $C_{2n}$ that
defines the particular thin disc model through (\ref{eq:poten}).

As we can see, the value of $C_0$ it is not determined by expression
(\ref{relation}). However, it is clear from (\ref{density}) that the surface
mass density diverges at the disc edge, when $\eta = 0$, unless that we impose
the condition (\citeauthor{HUN1} \citeyear{HUN1})
\begin{equation}
\sum_{n = 0}^{m} C_{2n} (2n+1) q_{2n+1}(0) P_{2n}(0) = 0, \label{finitden}
\end{equation}
that, after use the properties of the Legendre functions, leads to the
expression
\begin{equation}
C_{0} = \sum_{n = 1}^{m} (-1)^{n + 1} C_{2n}, \label{constant0}
\end{equation}
which gives the value of $C_0$, and then of the total mass $\mathcal{M}$, in terms of the
$A_{2n}$.

The previous expressions imply then that any particular thin disc model will be
completely determined by a set of constants $A_{2n}$, which must be choosed in
such a way that the circular velocities can be adjusted very accurately to fit
the observed rotation curves. Furthermore, as the powers ${\widetilde R}^{2n}$
are a set of linearly independents functions, the expression (\ref{velr2n}) is
quite adequate to be numerically adjusted to any set of data. Accordingly,
expression (\ref{velr2n}) can be considered as a kind of ``universal rotation
curve'' for flat galaxies, which can easily be adjusted to the observed data of
the rotation curve of any particular spiral galaxy.

Now then, besides the circular velocity, there are two other important
quantities concerning the kinematics of the models, which describe the stability
against radial and vertical perturbations of particles in quasi-circular orbits
(\citeauthor{BT} \citeyear{BT}). These two quantities, which must be positive in
order to have stable circular orbits, are  the epicyclic or radial frequency,
defined as
\begin{equation}
\kappa^{2} (R) = \left. \frac{\partial^{2} \Phi_{\rm eff}}{\partial R^{2}}
\right|_{z=0}, \label{epiciclic}
\end{equation}
and the vertical frequency, defined as
\begin{equation}
\nu^{2} (R) = \left. \frac{\partial^{2} \Phi_{\rm eff}}{\partial z^{2}}
\right|_{z=0}, \label{vertical}
\end{equation}
where
\begin{equation}
\Phi_{\rm eff} = \Phi (R,z) + \frac{\ell^{2}}{2R^{2}}, \label{phieff}
\end{equation}
is the effective potential and $\ell = R v_c$ is the specific axial angular
momentum.

By using then (\ref{vc2}) and (\ref{phieff}) in (\ref{epiciclic}), we can easily
obtain the relation
\begin{equation}
\kappa^2 (R) = \frac{1}{R} \frac{dv_c^2}{dR} + \frac{2 v_c^2}{R^2},
\end{equation}
in such a way that, by using (\ref{velr2n}), the epiciclic frequency can be cast
as
\begin{equation}
{\widetilde \kappa}^2 ({\widetilde R}) = \sum_{n = 1}^{m} 2 (n + 1) A_{2n}
{\widetilde R}^{2n - 2}, \label{kapr2n}
\end{equation}
where ${\widetilde \kappa} = a\kappa$. Also, from the expression for the Laplace
operator in cylindrical coordinates and using (\ref{vc2}), (\ref{vertical}) and
(\ref{phieff}), is easy to see that
\begin{equation}
\nu^2 (R) = \left. \nabla^2 \Phi \right|_{z=0} - \frac{1}{R} \frac{dv_c^2}{dR},
\label{lap0}
\end{equation}
so, as the potential is a solution of the Laplace equation, by using
(\ref{velr2n}) the vertical frequency can be written as
\begin{equation}
{\widetilde \nu}^2  ({\widetilde R}) = - \sum_{n = 1}^{m} 2 n A_{2n} {\widetilde
R}^{2n - 2}, \label{nur2n}
\end{equation}
where ${\widetilde \nu} = a \nu$.

\section{Fitting of Data to the Models}\label{sec:fit}

In order to adjust the previous model to real observed data, we choose four
spiral galaxies of the Ursa Major cluster, the galaxies NGC3877, NGC3917,
NGC3949 and NGC4010. The corresponding data are taken from the recent paper by
\cite{VS}, which presents the results of an extensive 21 cm-line synthesis
imaging survey of 43 galaxies in the nearby Ursa Major cluster using the
Westerbork Synthesis Radio Telescope. For each rotation curve data, we take the value of $a$ as given by the last
tabulated value of the radius. Accordingly, we are assuming that the radii of
the galaxies are defined by the last observed data. Then, by taking the radii
normalized in units of $a$, we make a non-linear least square fit of the data
with the general relation (\ref{velr2n}), by considering in each case a value of
$m$ less than the number of available data points.

\begin{figure}
$$\begin{array}{c}
\epsfig{width=3in,file=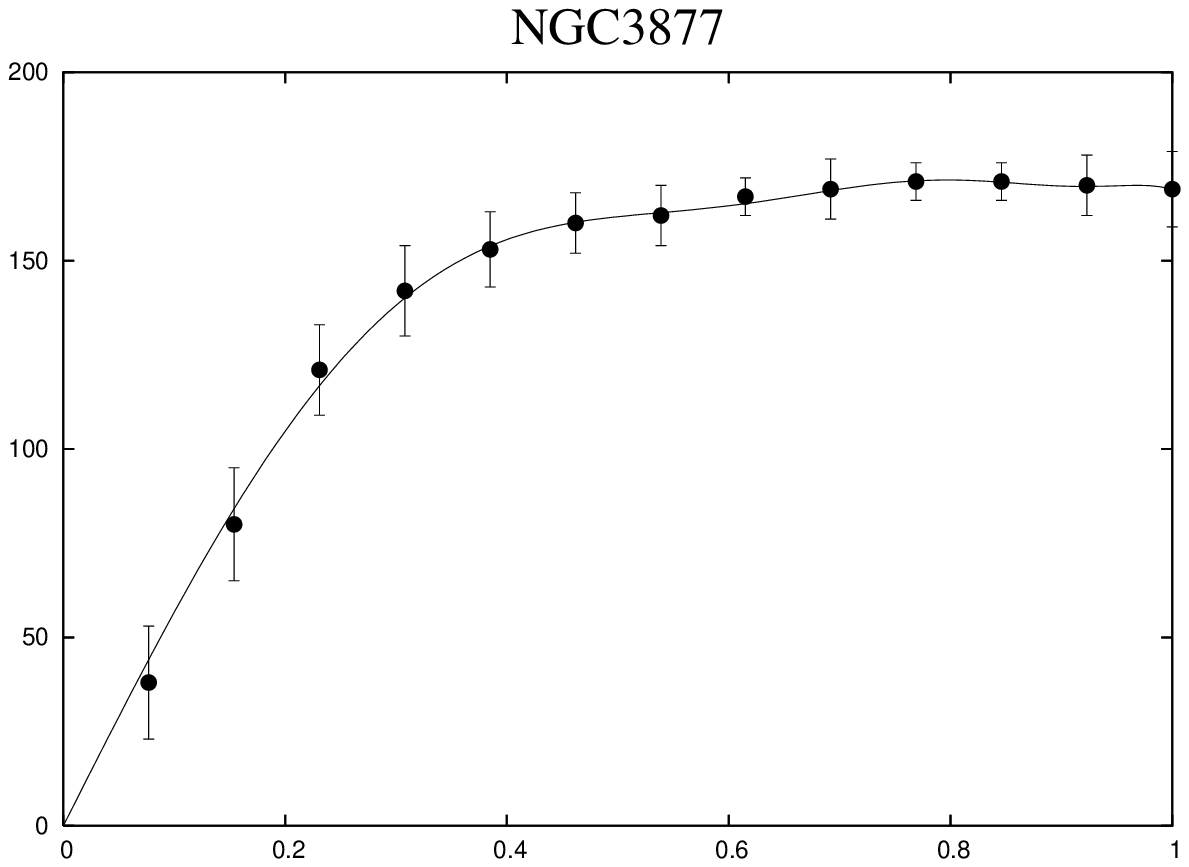} \\
\epsfig{width=3in,file=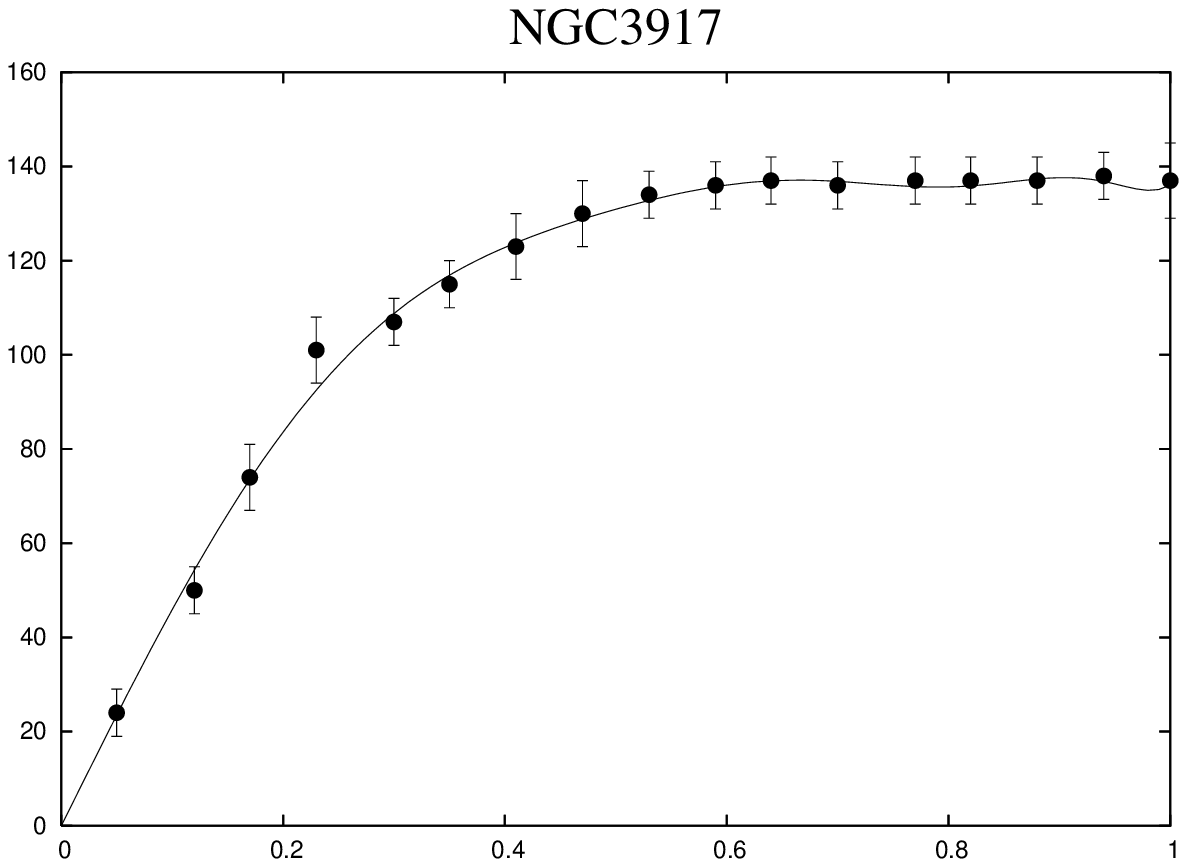} \\
\epsfig{width=3in,file=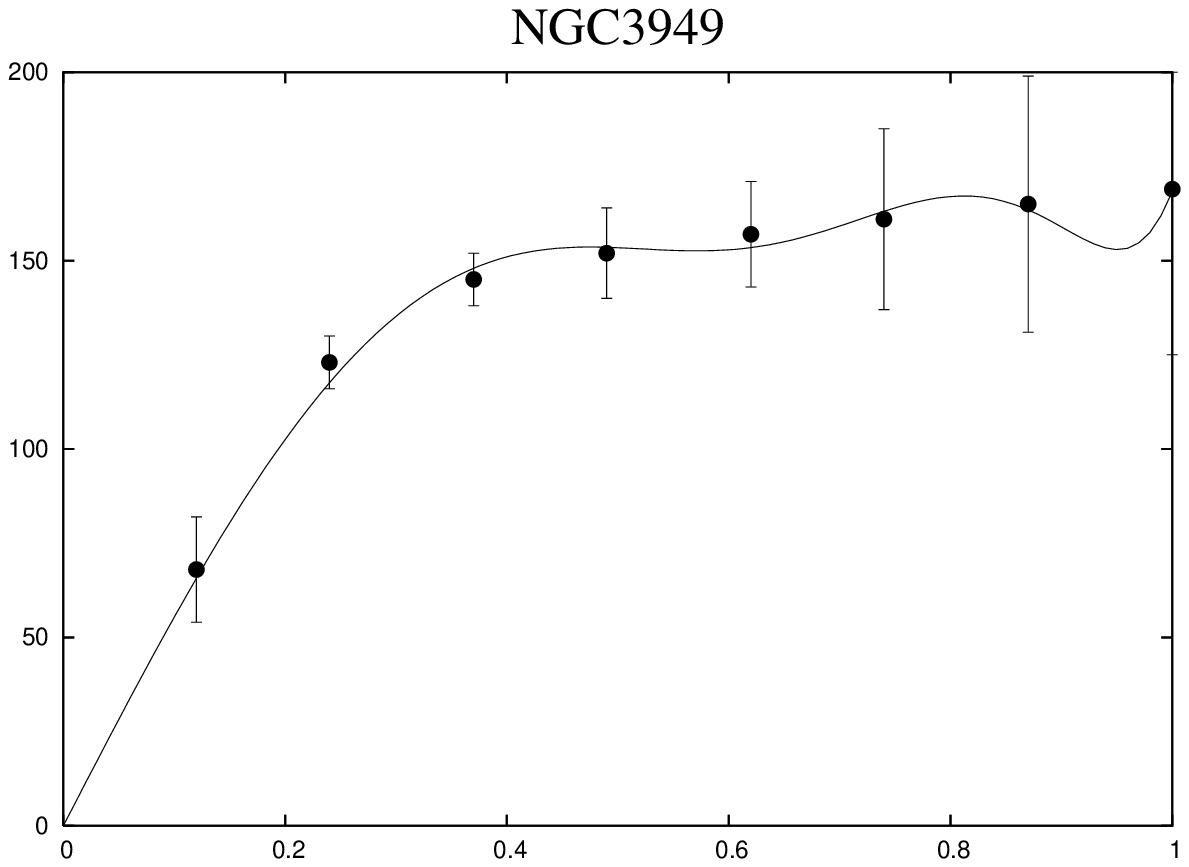} \\
\epsfig{width=3in,file=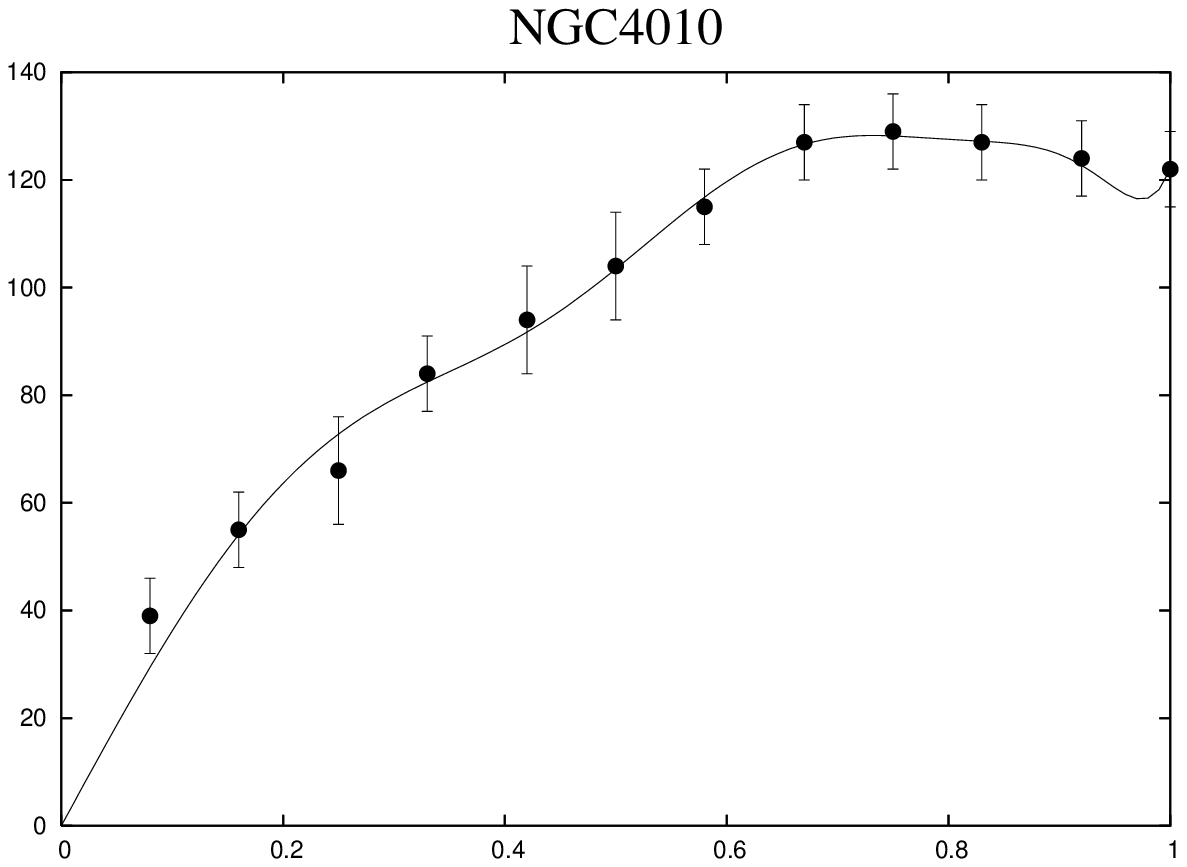} \\
\end{array}$$
\caption{Plots of the circular velocities $v_{c}$ in $\rm km/s$, as functions of
the dimensionless radial coordinate ${\widetilde R} = R/a$, for the spiral
galaxies NGC3877, NGC3917, NGC3949 and NGC4010.}\label{rotcurv}
\end{figure}

In Figure \ref{rotcurv} we show the adjusted rotation curves for the four
galaxies considered. The points with error bars are the observations, as
reported in \cite{VS}, while the solid line is the rotation curve determined
from (\ref{velr2n}) with the values for the $A_{2n}$ given by the numerical fit.
As we can see, the relation (\ref{velr2n}) fit quite accurately to the observed
data of the four galaxies considered. Then, from the obtained values for
$A_{2n}$, the corresponding values of the $C_{2n}$ are determined by using
(\ref{relation}) and (\ref{constant0}). In Table \ref{tab:c2n} we present the
values of $C_{2n}$ for the four galaxies as well as the corresponding value of
$m$ used in (\ref{velr2n}). In Table \ref{tab:mass} we indicate, for each
galaxy, the morphological type according to the Hubble's classification of
galaxies, the radius $a$ in ${\rm kpc}$ and the total mass $\mathcal{M}$, both
in ${\rm kg}$ and in solar mass units ($\mathcal{M_{\odot}}$).

\begin{table}
\caption{Constants $C_{2n}$ $[{\rm km}^2 {\rm s}^{-2}]$ and values of
$m$.}\label{tab:c2n}
\begin{tabular}{lrrrr}
\hline
	& NGC3877 & NGC3917 & NGC3949 & NGC4010 \\ \hline
$m$   &	$6$ & $7$ & $5$	& $7$	\\ 
$C_0$ &	17452.78 & 11258.92 & 15686.77 & 8801.12 \\ 
$C_2$ &	26564.93 & 17210.85 & 24051.95 & 12875.61 \\ 
$C_4$ &	13926.79 & 8998.92 & 13011.66 & 4944.66 \\ 
$C_6$ &	7478.82 & 4573.34 & 7127.38 & 1117.08 \\ 
$C_8$ &	4053.13 & 2213.64 & 4577.68 & 866.64 \\ 
$C_{10}$ & 1887.23 & 1063.01 & 2096.77 & 1019.52 \\ 
$C_{12}$ & 498.29 & 640.40 &	& 818.98 \\ 
$C_{14}$ &	& 264.69 &	& 419.19 \\ \hline
\end{tabular}
\end{table}

\begin{table}
\caption{Morphological type, radius $a$ and total mass
$\mathcal{M}$.}\label{tab:mass}
\begin{tabular}{llccc}
\hline
& Type & $a$ $[\rm{kpc}]$ & $\mathcal{M}$ $[{\rm kg}]$ & $\mathcal{M}$
$[\mathcal{M_{\odot}}]$  \\ \hline
NGC3877 & Sc & 11.74 & 9.47$\ \times \ 10^{40}$ & 4.76$\ \times \ 10^{10}$ \\ 
NGC3917 & Scd & 15.28 & 7.95$\ \times \ 10^{40}$ & 3.95$\ \times \ 10^{10}$ \\ 
NGC3949 & Sbc &\  8.72 & 6.32$\ \times \ 10^{40}$ & 3.18$\ \times \ 10^{10}$ \\ 
NGC4010 & SBd & 10.84 & 4.41$\ \times \ 10^{40}$ & 2.22$\ \times \ 10^{10}$ \\
\hline
\end{tabular}
\end{table}

\begin{figure}
$$\begin{array}{c}
\epsfig{width=3in,file=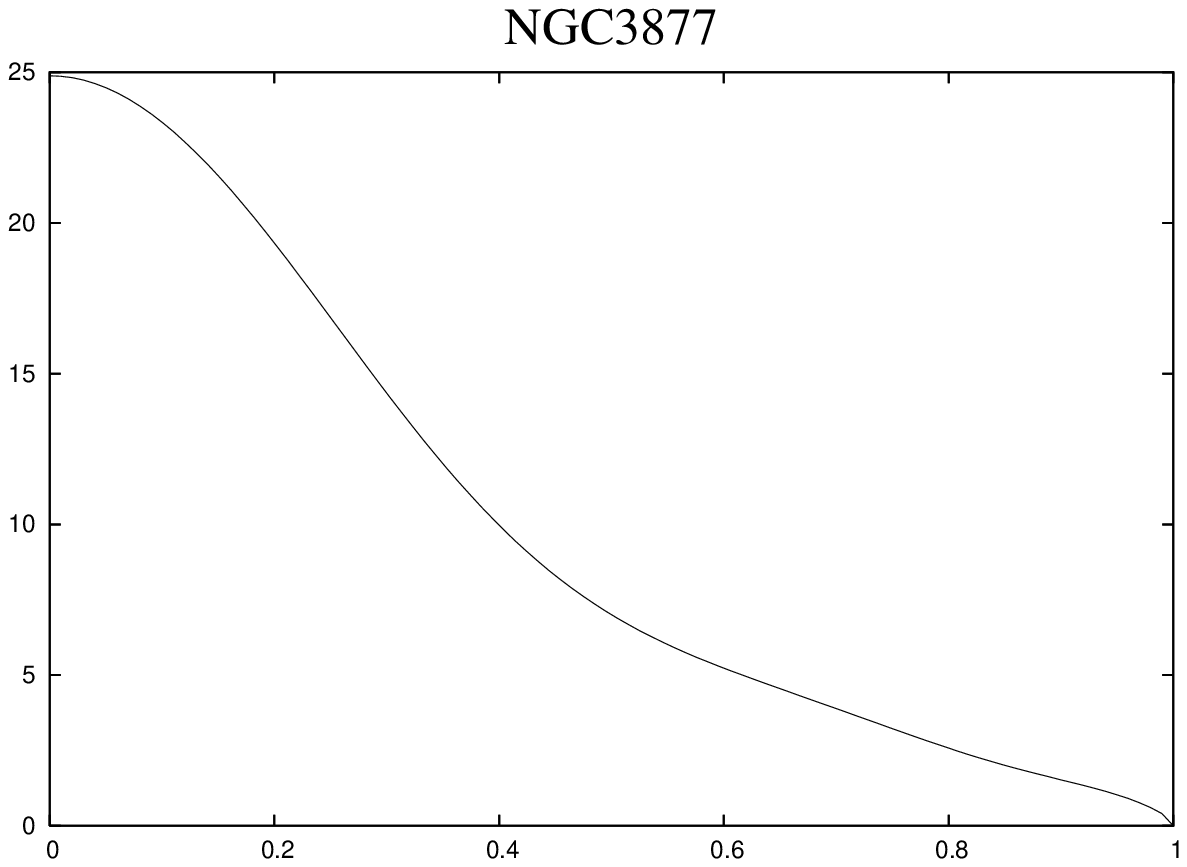} \\ 
\epsfig{width=3in,file=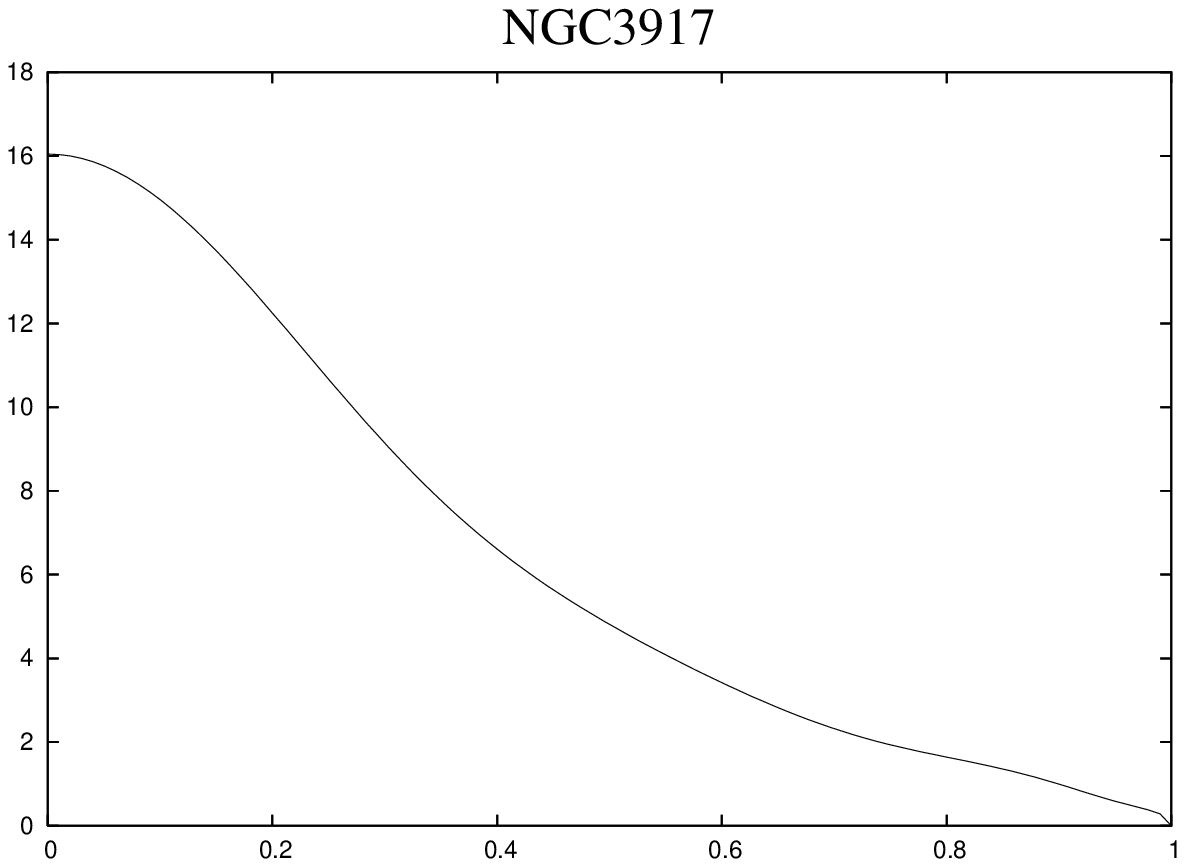} \\
\epsfig{width=3in,file=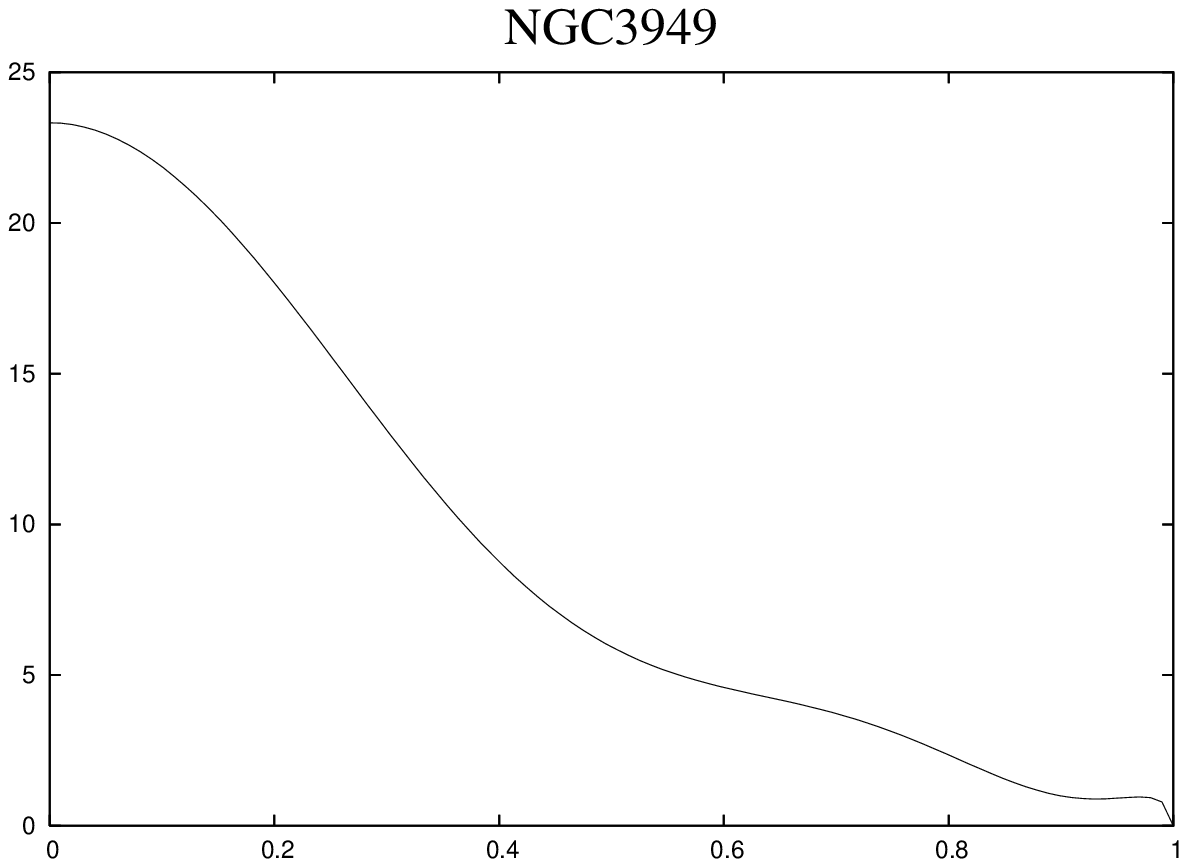} \\
\epsfig{width=3in,file=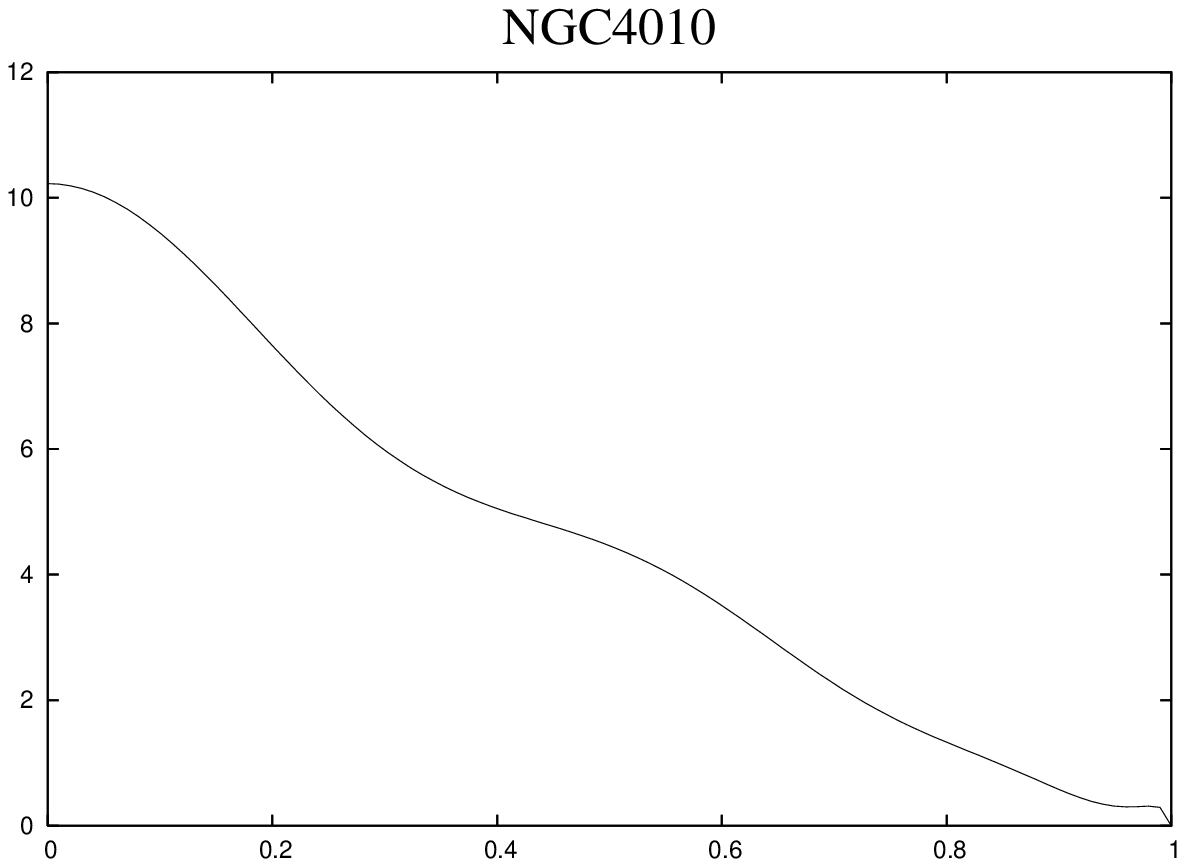} \\
\end{array}$$
\caption{Plots of the surface densities $\Sigma \times 10^{-3}$ in $\rm kg/m^2$,
as functions of the dimensionless radial coordinate ${\widetilde R} = R/a$, for
the spiral galaxies NGC3877, NGC3917, NGC3949 and NGC4010.}\label{densit}
\end{figure}

\begin{figure}
$$\begin{array}{c}
\epsfig{width=3in,file=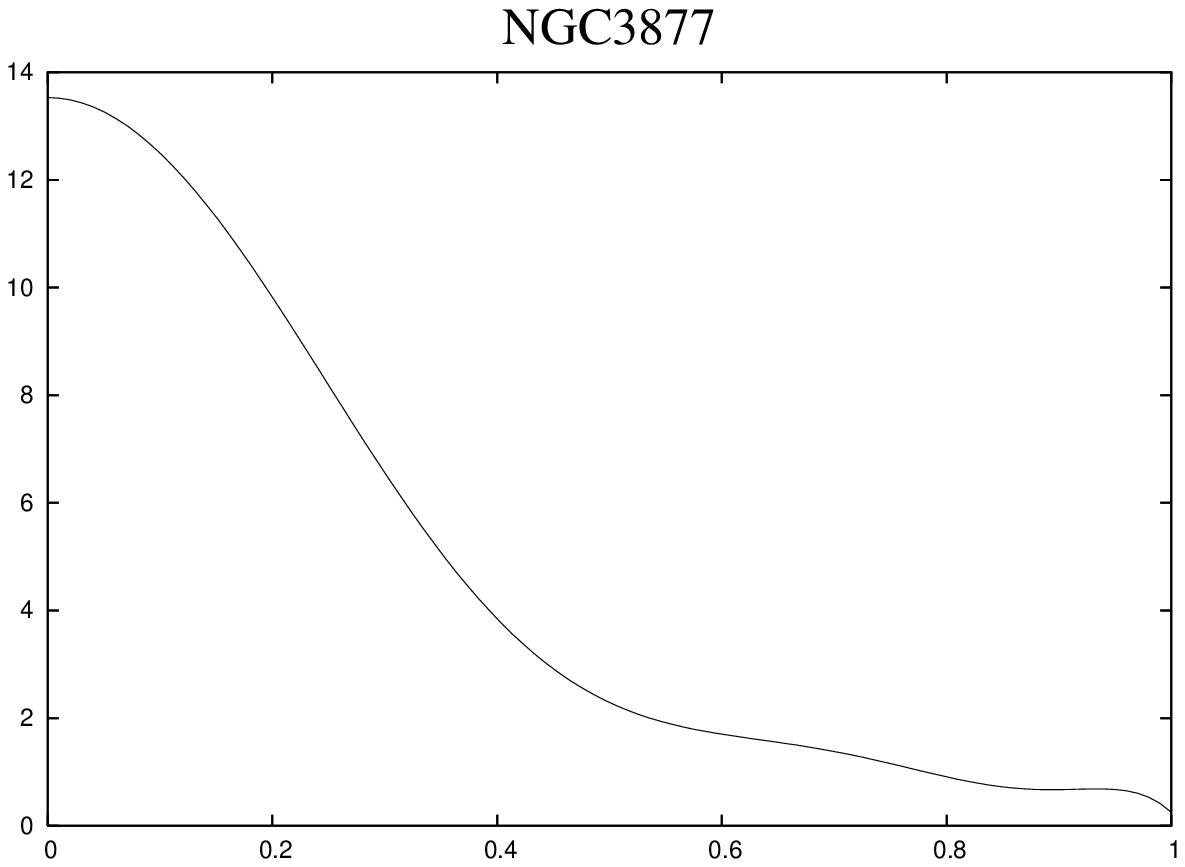} \\
\epsfig{width=3in,file=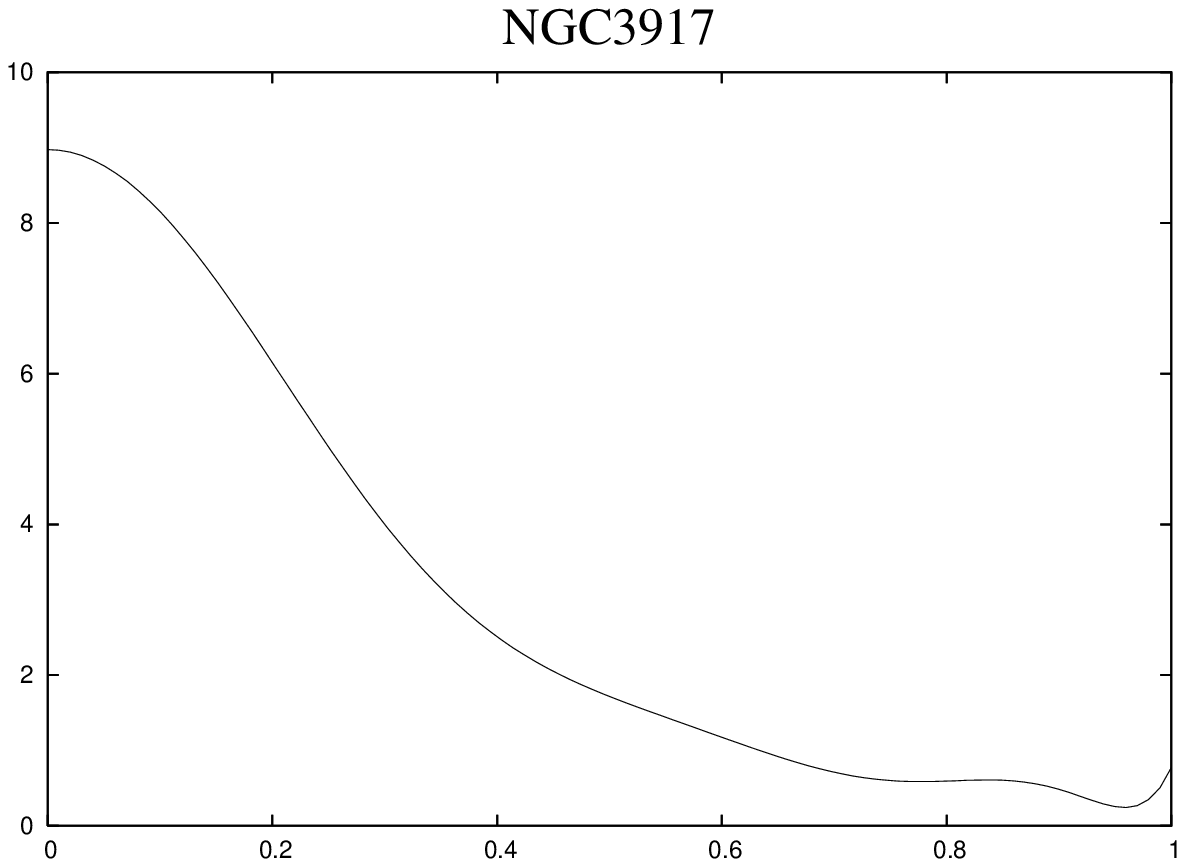} \\
\epsfig{width=3in,file=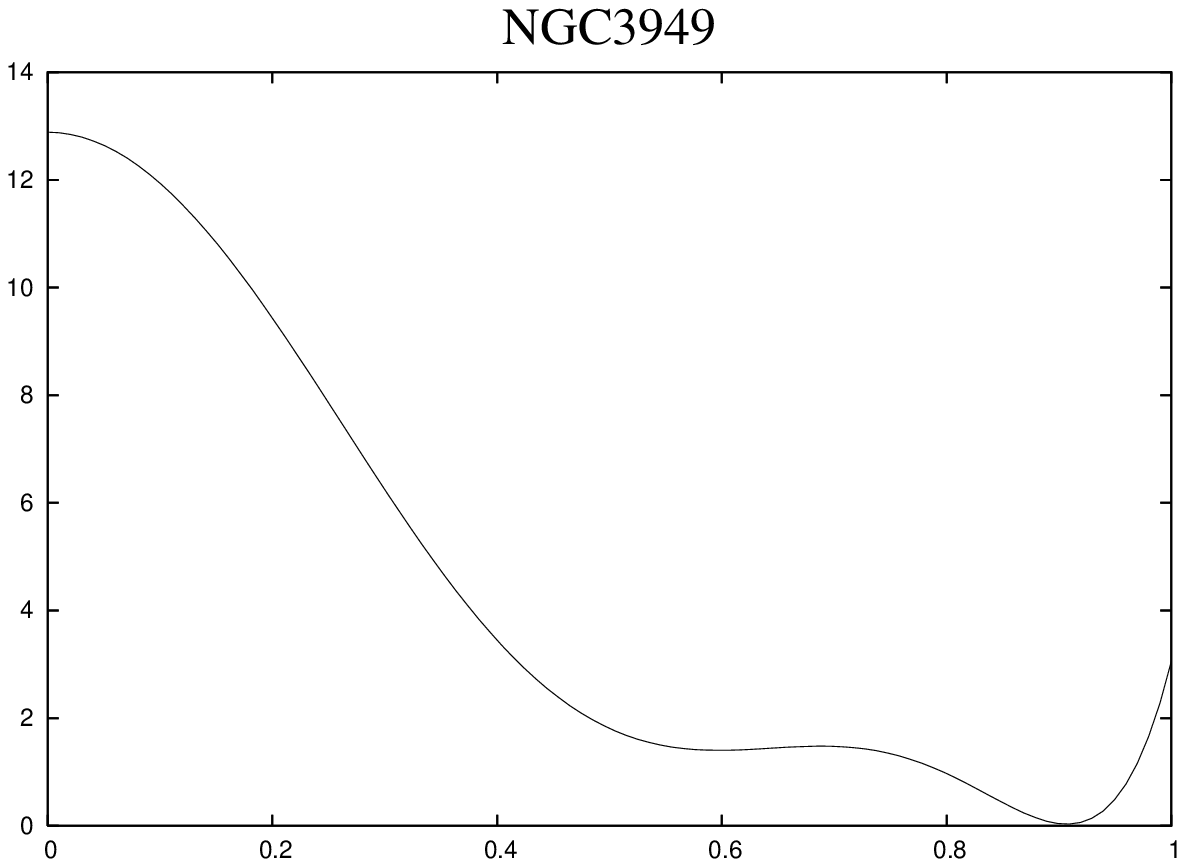} \\
\epsfig{width=3in,file=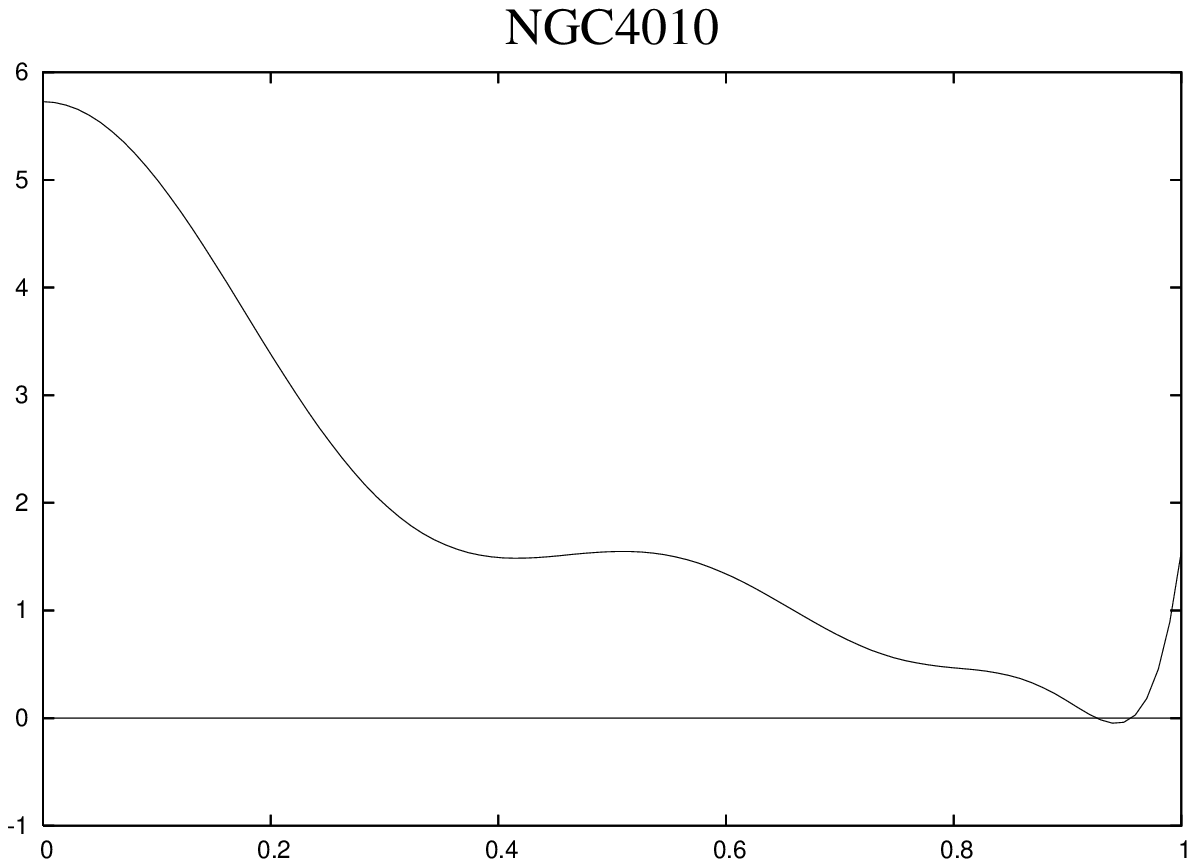} \\
\end{array}$$
\caption{Plots of the epiciclic frequencies ${\widetilde \kappa}^2 \times
10^{-5}$ in $({\rm km/s})^{2}$, as functions of the dimensionless radial
coordinate ${\widetilde R} = R/a$, for the spiral galaxies NGC3877, NGC3917,
NGC3949 and NGC4010.}\label{epicic}
\end{figure}

\begin{figure}
$$\begin{array}{c}
\epsfig{width=3in,file=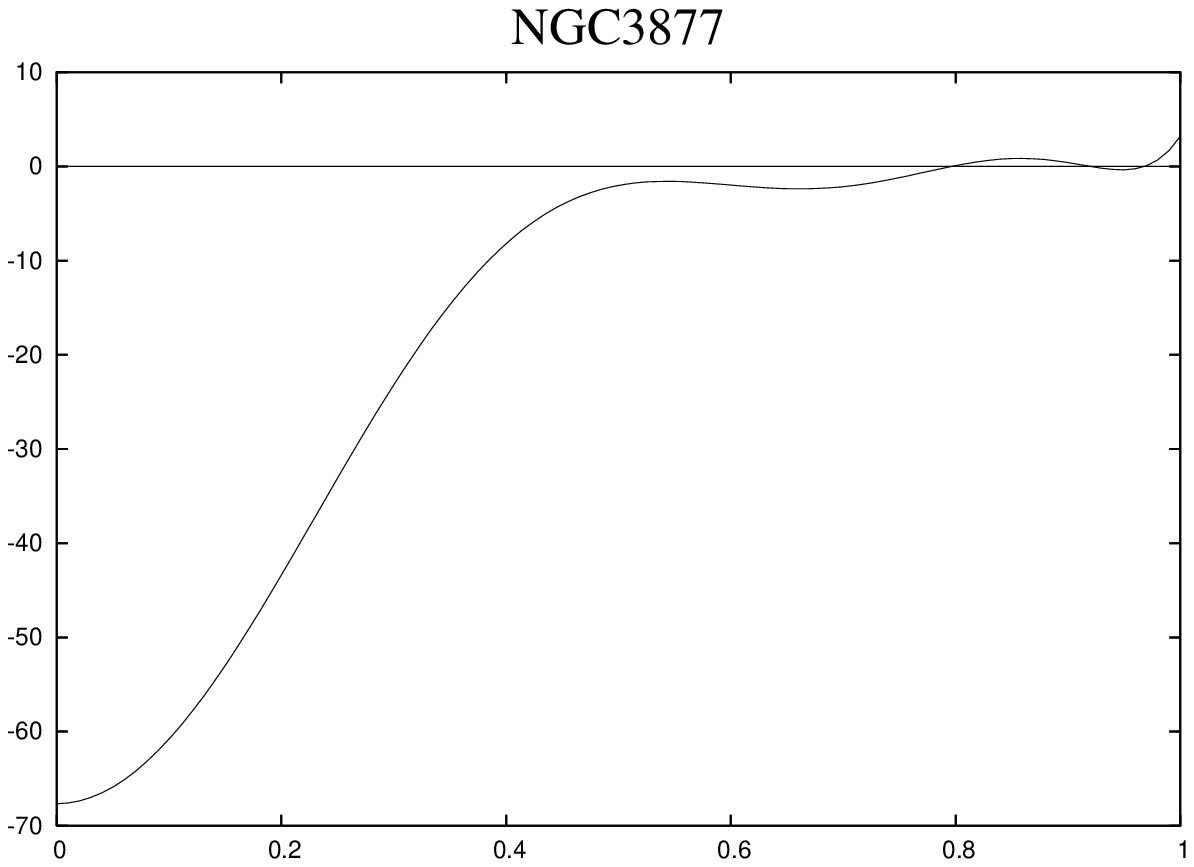} \\
\epsfig{width=3in,file=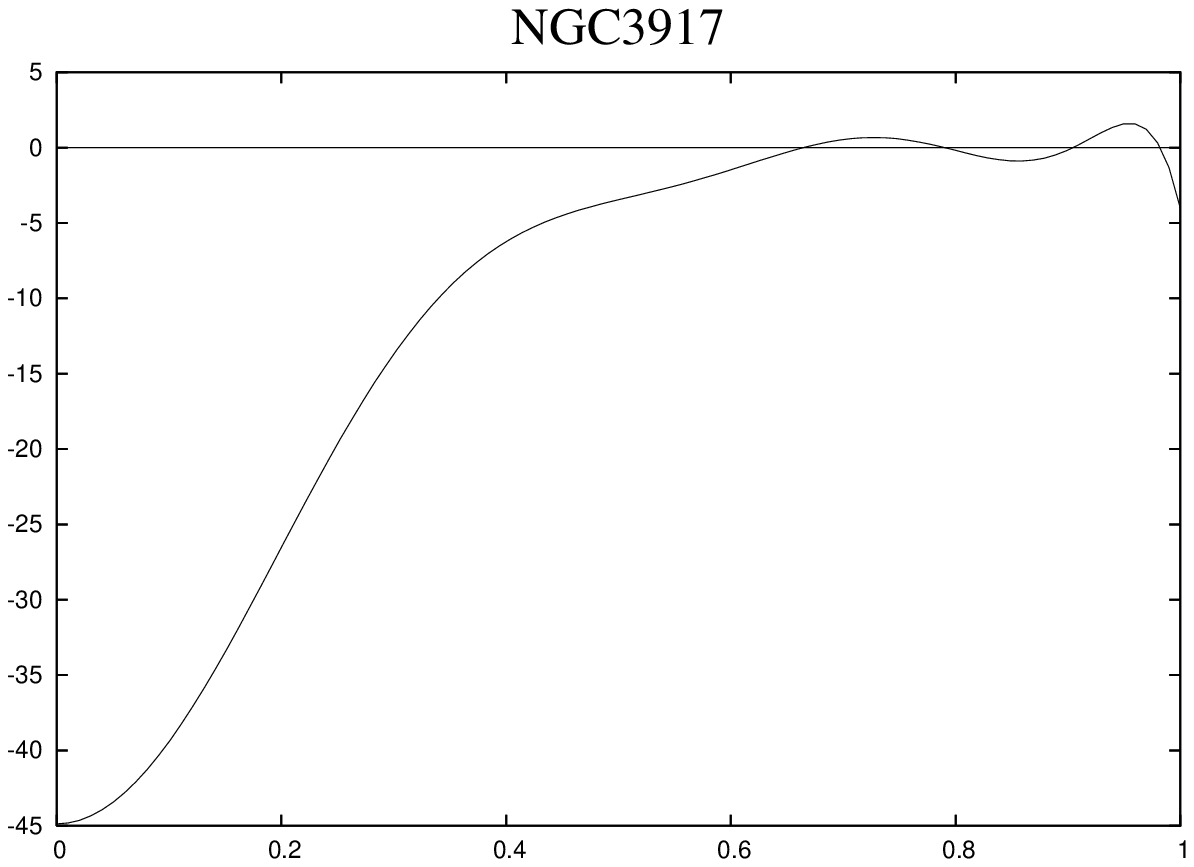} \\
\epsfig{width=3in,file=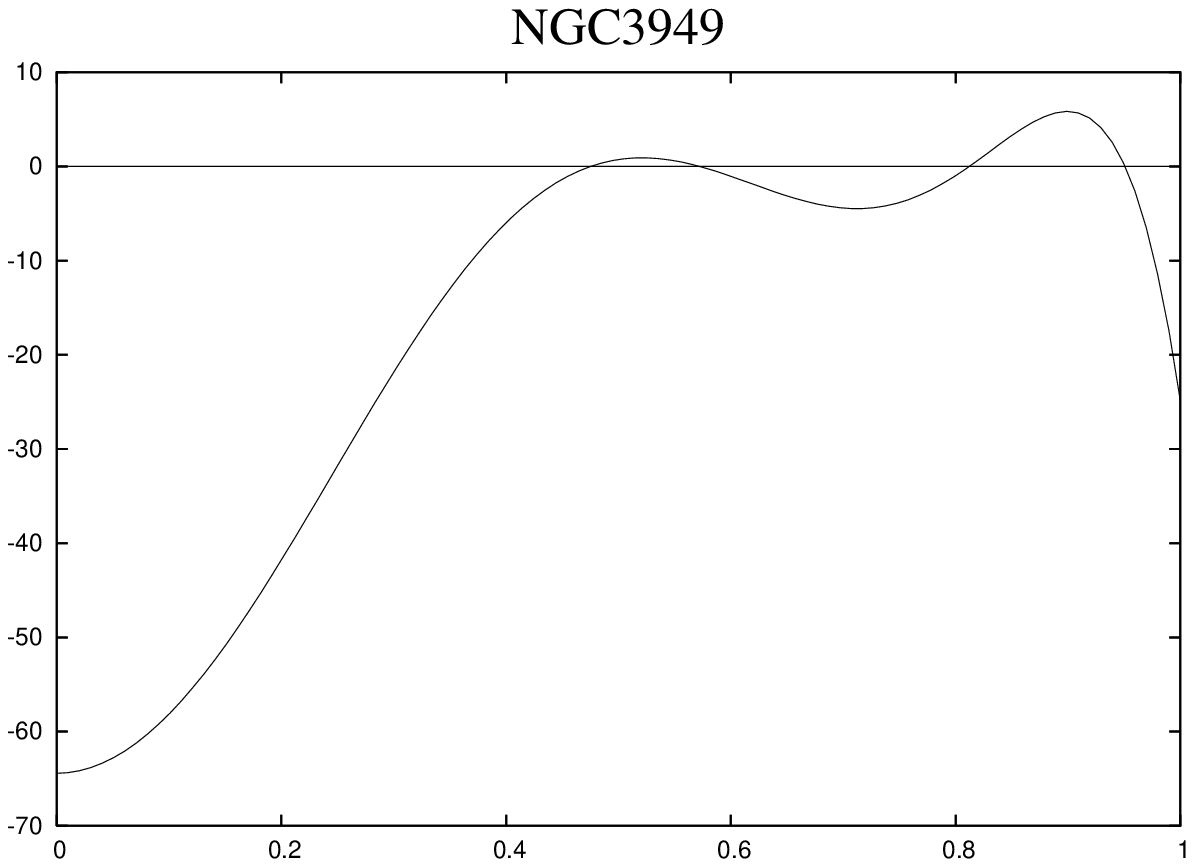} \\
\epsfig{width=3in,file=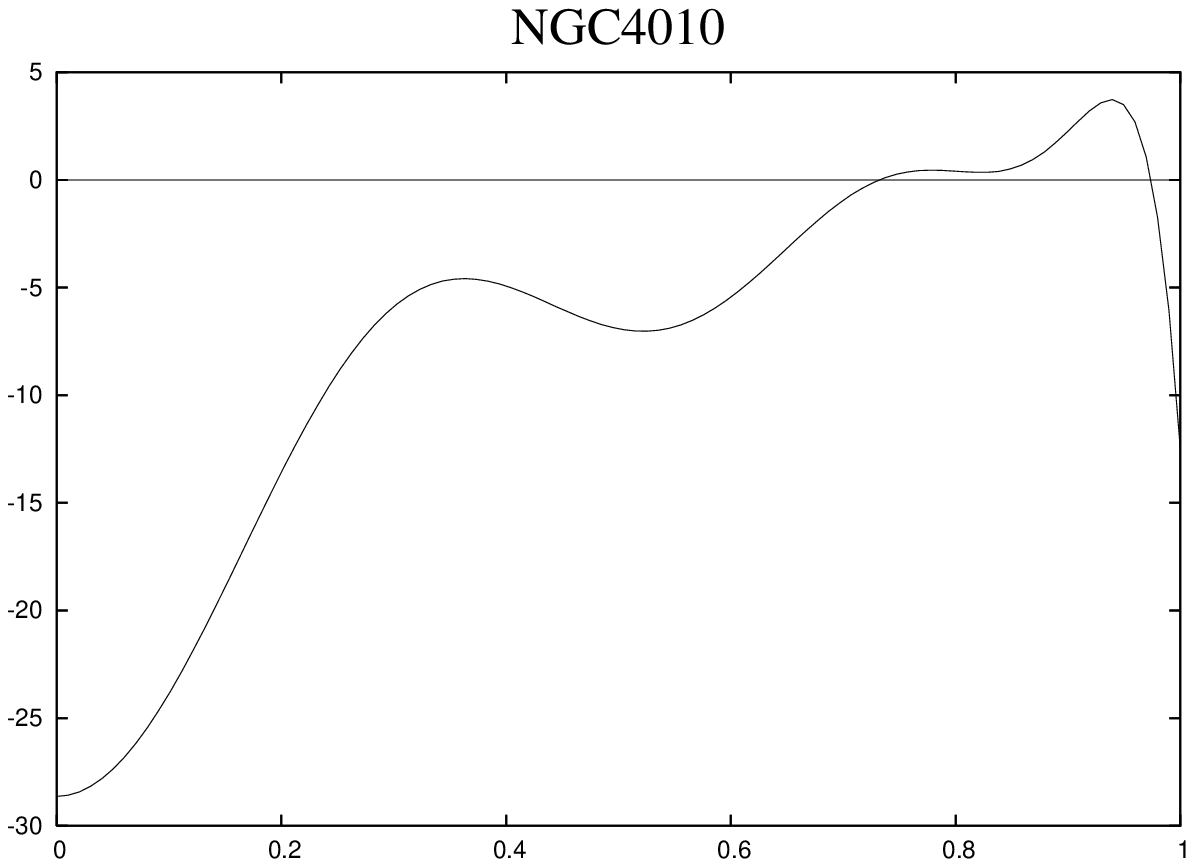} \\
\end{array}$$
\caption{Plots of the vertical frequencies ${\widetilde \nu}^2 \times 10^{-4}$
in $({\rm km/s})^{2}$, as functions of the dimensionless radial coordinate
${\widetilde R} = R/a$, for the spiral galaxies NGC3877, NGC3917, NGC3949 and
NGC4010.}\label{vertic}
\end{figure}

Now, as the set of constants $C_{2n}$ it defines completely each particular thin
disc model, we can easily compute all the physical quantities characterizing
each galaxy. However, as explicit expressions for the gravitational potential
$\Phi (R,z)$ and the surface mass density $\Sigma (R)$ can be easily obtained by
using the values of the $C_{2n}$ at expressions (\ref{eq:poten}) and
(\ref{density}), we will not present them here. Insteed, we plot in Figure
\ref{densit} the surface densities for the four galaxies, as functions of the
dimensionless radial coordinate ${\widetilde R} = R/a$. For the four galaxies we
obtain a well behaved surface mass density, having a maximum at the disc center
and then decreasing until vanish at the disc edge.

In a similar way, we can compute the epiciclic and vertical frequencies by using
(\ref{kapr2n}), (\ref{nur2n}) and the values of the constants $A_{2n}$ obtained
from the numerical fit. However, as with the surface mass densities, we will not
present the explicit expressions here and, instead, we only show the
corresponding plots. So, in Figure \ref{epicic}, we show the plots of the
epiciclic frequencies for the four galaxies considered and, in Figure
\ref{vertic}, the corresponding plots of the vertical frequencies. From the
plots at Figure \ref{epicic} we can see that only the galaxy NGC4010 presents a
small region of radial instability near the disc edge. On the other hand, as it
is shown at Figure \ref{vertic}, the four galaxies are instable against vertical
perturbations.

\section{Discussion}\label{sec:disc}

We presented four particular thin disc models adjusted in order to accurately
fit the observed data of the rotation curves for the galaxies NGC3877, NGC3917,
NGC3949 and NGC4010 of the Ursa Major cluster. These models present well behaved
surface densities, that resembles the observed luminosity profile of many spiral
galaxies, and the obtained values for the corresponding total mass $\mathcal{M}$
it agrees with the expected order of magnitude. Accordingly, the here obtained
expression for the circular velocity, equation (\ref{velr2n}), can be considered
as a kind of ``universal rotation curve'' for flat galaxies, which can easily be
adjusted to the observed data of the rotation curve of any particular spiral
galaxy.

On the other hand, in one of the models we obtain a small region near the disc
edge with instability agains radial perturbations. Now, as the models are
completely determined by the set of constants $A_{2n}$, which are fixed by the
numerical fit of the rotation curve data, there are not free parameters that can
be adjusted by requiring radial stability. A posible solution for this problem
may be consider a less restrictive numerical fit that leaves some free
parameters. This can be done by taking the summation in expression 
(\ref{velr2n}) until a value of $m$ greater than the available number of data
points.

However, the models present a central region with strong instabilities against
vertical perturbations of particles in quasi-circular orbits. This result was
expected as a consequence of the fact that the models only consider the thin
galactic disc. Indeed, as we can see from expression (\ref{lap0}), vertical
instability will be always present in models constructed from solutions of
Laplace equation and adjusted in such a way that their circular velocities
reproduce the central region of the observed rotation curves, where the velocity
rises linearly with the radius. Therefore, more realistic models must be
considered that include the non-thin character of the galactic disc, or the mass
contribution of the spheroidal halo.

In agreement with the above considerations, we can consider the set of models
here presented as a first approximation to the obtaining of quite realistic
models of spiral galaxies. In particular, we believe that the values of
$\mathcal{M}$ that were obtained for the four galaxies studied may be taken as a
quite accurately estimative of the mass upper bound of these galaxies, since in
the model was considered that all their mass was concentrated at the galactic
disc. Accordingly, we are working now in a generalization of the model that
includes the mass contribution of the spheroidal halo, in such a way that we can
overcome the vertical instability problem and obtain some estimative of the
relative contributions of the halo and the disc to the total mass of the
galaxies.

\end{document}